# Guidelines for the implementation of power oscillation damping controllers in power converters

**Javier RENEDO***
**Red Eléctrica - Redeia**
**Spain**
**javier.renedo@ree.es**

**Macarena MARTÍN ALMENTA**
**Red Eléctrica - Redeia**
**Spain**
**macarena.martin@ree.es**

**Sergio MARTÍNEZ VILLANUEVA**
**Red Eléctrica - Redeia**
**Spain**
**smartinez@ree.es**

**Agustín DÍAZ-GARCÍA**
**Red Eléctrica - Redeia**
**Spain**
**agustin.diaz@ree.es**

**Antonio CORDÓN**
**Red Eléctrica - Redeia**
**Spain**
**acordon@ree.es**

**Davide GOTTI**
**Red Eléctrica - Redeia**
**Spain**
**davide.gotti@ree.es**

## SUMMARY

One of the most effective ways to damp electromechanical oscillations in power systems is by means of supplementary controllers attached to the different devices in the power system is by means of power system stabilizers (PSS) in synchronous machines or by means of power oscillation damping (POD) controllers in facilities with power converters. In the recent years, the use of POD controllers in voltage source converters (VSCs) with grid-following (GFL) control has been investigated. Although the potential of POD controllers to help to damp inter-area oscillations in power systems is enormous, their correct implementation is not trivial, because their effectiveness is strongly linked to their settings. This paper provides guidelines for the implementation of POD controllers in power converters for application in real-world power systems. The paper proposes compliance criteria for POD controllers using a synthetic test system and a systematic methodology used in Spanish technical standard for monitoring compliance (NTS), considering practical considerations. The paper also includes numerical examples to illustrate compliance criteria for POD controllers in a synthetic test system. A generic power converter with grid-following (GFL) control is used for the analysis by simulation and POD controllers using modulation of active-power injection (POD-P), reactive-power injection (POD-Q) or both simultaneously (POD-PQ) will be analysed. Results were validated in a large-scale power system. The paper concludes that by using appropriate synthetic systems, methodologies and compliance criteria, POD controllers in power converters could be effective to damp electromechanical oscillation in large-scale power systems.

# 1 Introduction

Electromechanical oscillations are related to angle stability under small disturbances, and they involve generators that oscillate against each other through the power system. Electromechanical oscillations can be divided into local and inter-area oscillations [1]. Local oscillations involve oscillations between groups of coherent generators located close to each other, with oscillation frequencies of these modes ranging from 0.7 to 2.5 Hz. Inter-area oscillations involve oscillations between groups of coherent generators in different areas which are located far from each other, and their oscillation frequencies can be from 0.1 to 1 Hz.

One of the most effective ways to damp electromechanical oscillations in power systems is by means of supplementary controllers attached to the different devices in the power system:

- Synchronous machines: by means of power system stabilizers (PSS).
- Facilities containing power converters, such as power park modules (PPM) (e. g., solar photovoltaic (PV) plants, wind power plants, etc…), energy storage systems (ESS), Flexible Alternating Current Transmission Systems (FACTS), or high voltage direct current (HVDC) power transmission systems: by means of power oscillation damping (POD) controllers.

In the past years, the use of POD controllers in voltage source converters (VSCs) with grid-following (GFL) control has been investigated. For example, some research publications have studied POD controllers applied to wind power parks [2], [3], solar photovoltaic (PV) power parks [4], power converters in general [5], FACTS systems [6] and VSC-HVDC systems [7], [8], [9]. Recently, some publications have analysed POD controllers in grid-forming (GFM) power converters [10]. POD controllers have also been implemented successfully in real-world facilities during the past years. For example, reference [11] describes the implementation of a POD controller in a wind power park, references [12], [13] describe the practical implementation of POD controllers in FACTS devices and the work in [14] describes tests on the POD-P controller of INELFE-1 VSC-HVDC Spain-France interconnector.

Although the potential of POD controllers to help to damp inter-area oscillations in power systems is enormous, their correct implementation is not trivial, because their performance is strongly linked to their settings. POD controllers can be implemented in TSO facilities (FACTS, HVDC systems, etc..) or in facilities owned by other stakeholders (e.g., PPMs, etc..). In the latter, the implementation of POD controllers with proper settings may be more challenging, because the owner of the facility or the manufacturer of power converters typically do not have available detailed models of the large-scale power system. Some of the difficulties in implementation of POD controllers are illustrated in [3] and [15], where aspects related to POD input signals and coexistence of POD-Q controllers with voltage control were discussed. To tackle the obstacles in the practical implementation of POD controllers, some Transmission System Operators (TSO) use synthetic systems to evaluate the performance of POD controllers with compliance simulations [16], [17], [18].

The proposed revision of the European network code for technical requirements for generators (EU 2016/631 (NC RfG) [19]), submitted by ACER for public consultation in July 2023, proposes technical requirements for POD controllers. The draft document is known as *Network*

*Code - Requirements for Generators 2 (NC RfG2)* [20]. At Spanish national level, the draft proposal of the Royal Decree for technical requirements for facilities connected to the transmission and distribution system [21] proposed technical requirements for POD controllers (mandatory), in line with the current European regulation [19], [22] and the Spanish regulation [23]. These requirements are general in nature and do not impose specific variants of POD controllers. Regardless of the freedom in the form of a particular POD controller, it is essential that the proposed POD controllers are aligned with the expected technical capabilities, so they can contribute to damp electromechanical oscillations in the power system.

The Spanish TSO launched a working group at national level (from June 2023 to June 2024) among the various stakeholders in the Spanish electricity energy sector to write a guide for the technical functionalities of POD controllers and compliance simulations. The objective of this working group was to develop technical specifications of POD controllers, guidelines for their implementation and compliance criteria (non-mandatory technical specifications). The working group produced a guide for the implementation of POD controllers [18].

This paper provides guidelines for the implementation of POD controllers in power converters for application in real-world power systems, based on the main guidelines learnt from the working group. The paper also includes numerical examples to illustrate compliance criteria for POD controllers, using the synthetic two-area system used in Spanish technical standard for monitoring compliance (NTS) [24]. A generic power converter is used for the analysis by simulation and POD controllers using modulation of active-power injection (POD-P), reactive-power injection (POD-Q) or both simultaneously (POD-PQ) will be studied. Results were validated in a large-scale power system. The objective of analysing the performance of POD controllers in a large-scale power system is to answer the following question:

- *If a POD controller in a power converter is tuned using appropriate synthetic systems, methodologies and compliance criteria, is it still effective when the power converter is connected to a large-scale power system?*

This question has a great interest from a practical point of view, especially when using POD controllers in power converters when limited information of the power system is available. Naturally, the answer to the question is not trivial and the key is to use an appropriate synthetic system and methodology to assess the performance of the POD controller. These aspects will be discussed in the paper.

GFL technology is mature, and they have been used during the past decades. On the contrary, GFM technology is more recent, and its definition and technical requirements are currently being developed worldwide. For this reason, this paper just focuses on POD controllers in GFL power converters.

## 2 POD controllers in power converters

Figure 1-(a) shows a generic diagram of a voltage source converter (VSC) with GFL control connected to the power system, where $P$ and $Q$ represent the active- and reactive-power injections, respectively, $\bar{V} = V\angle\theta$ represents the voltage at the connection point, $\bar{I}$ represents the phasor of the current injection and $\omega$ represents the frequency at the connection point. The GFL device can represent a Power Park Module (PPM), an Electricity Storage Module of type PPM (ESM-PPM) or a shunt power converter. The facility will be referred in a generic way as *GFL device*.

For simplification purposes, it is assumed that the GFL device is controlled as:

- Active-power (P) control.
- Reactive-power (Q) control.

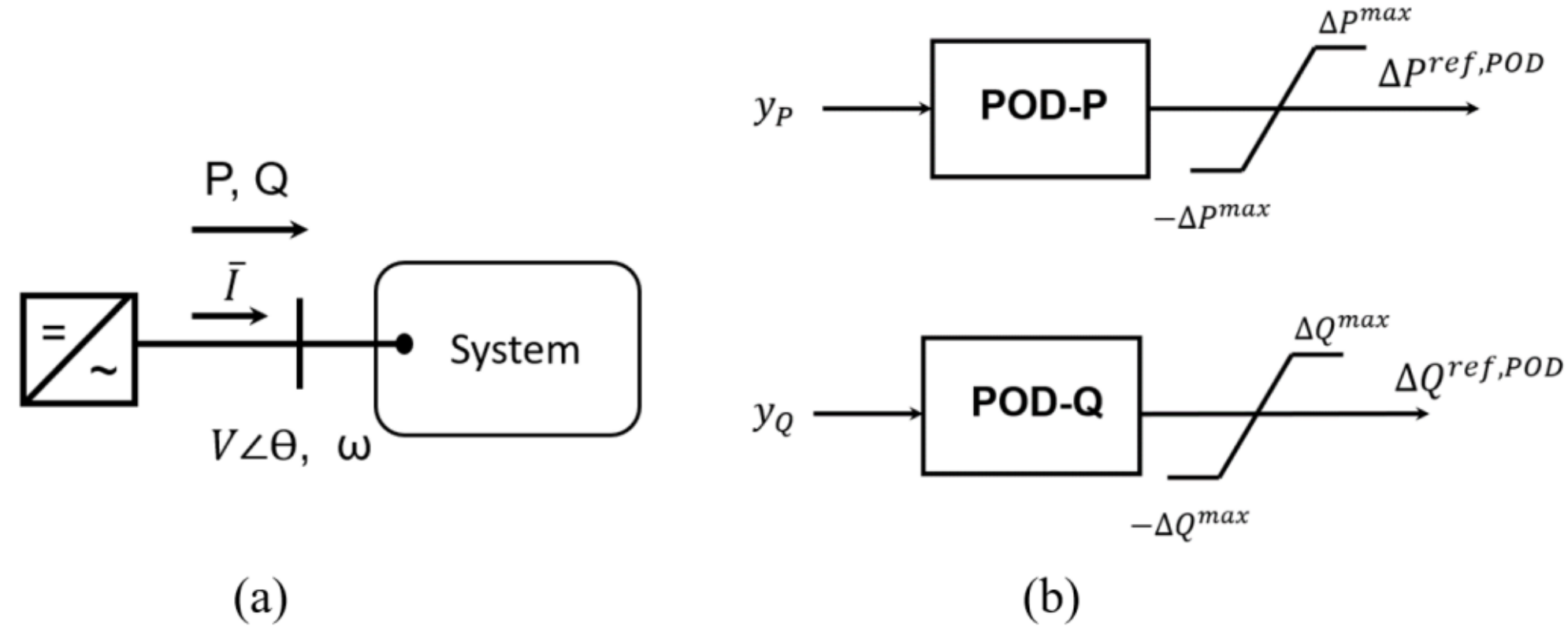

***Figure 1: (a) GFL device connected to the power system and (b) general scheme of POD controllers.***

The GFL device is equipped with POD controllers based on active-power modulation (POD-P) and reactive-power modulation (POD-Q). Active- and reactive-power setpoint values are given by:

$$P^* = P_0 + \Delta P^{ref,POD}, \quad Q^* = Q_0 + \Delta Q^{ref,POD}. \tag{1}$$

where:

- $P^*$ is the total active-power setpoint, $P^0$ is a constant active-power setpoint and $\Delta P^{ref,POD}$ is the supplementary active-power setpoint provided by POD-P controller.
- $Q^*$ is the total reactive-power setpoint, $Q^0$ is a constant reactive-power setpoint and $\Delta Q^{ref,POD}$ is the supplementary reactive-power setpoint provided by POD-Q controller.

Figure 1-(b) shows a generic block diagram of POD-P and POD-Q controllers of the GFL device. Each of them has an input signal, a generic block diagram, a saturator and the output signal is the supplementary P/Q setpoint provided by the POD controllers.

POD-P/POD-Q controllers could have different implementations:

- Different block diagrams.
- POD controllers could be implemented at Power Plant Controller (PPC) level or at converter level.
- Input signals: frequency at the connection point, voltage at the connection point, other local signals or remote signals.
- Output signals:
  - POD-P: supplementary reference for the active-power injection, direct-axis current, etc…
  - POD-Q: supplementary reference for reactive-power injection, transverse-axis current, or voltage, etc...
- Fixed/adaptive parameters.

The guide for the implementation of power oscillation damping controllers [18] discusses these aspects and it also presents most extended implementations. Stakeholders and manufacturers should have freedom to choose the implementation of POD controllers the design method to determine their settings, as long as they contribute to damp electromechanical oscillations in the power system in a reliable way. Therefore, it is important to define the expected technical capabilities of POD controllers and methods to evaluate systematically their performance [18].

## 3 Evaluation of the performance of POD controllers

POD controllers in PPMs, PPM-ESMs or any other shunt power converter facility are aimed to damp electromechanical oscillations in the power system. This is not trivial, because (a) the performance of POD controllers is strongly related to their settings and (b) the owner of the plants or manufacturers of power converters typically do not have available detailed models of the large-scale power system. A practical way to evaluate the performance of PSSs in synchronous machines and POD controllers in power converters, when limited information of the power system is available, is the use of appropriate synthetic test systems [25], [26], [27].

Reference [26] proposed a synthetic two-area test system and methodology to evaluate the performance of PSSs in synchronous machines when limited information of the power system is available. This synthetic test system and methodology were included in the Spanish Compliance Monitoring Technical Standard (NTS) [24]. The NTS standard also extended the application of this test system to evaluate the contribution of PPMs to inter-area-oscillation damping. The guide for the implementation of POD controllers [18] proposed the use of this synthetic test system and methodology to evaluate the performance of POD controllers.

Figure 2 shows the synthetic two-area test system [24], where the facility to be evaluated (GFL device in this case) is connected to bus 5. Notice that, according to the NTS methodology [24], the rating of the GFL device is normalised to 1500 MVA when connected to the synthetic two-area system, independently of its true rating. Electromechanical oscillations with different frequencies (0.1-1 Hz approximately) are reproduced by changing the line reactance ($X_L =$ $X_{23} = 0.01 - 0.6$ pu). This synthetic test system is very useful to analyse the impact of a facility connected to the Iberian Peninsula power system to the damping of inter-area oscillations in the CE power system. This test system will be used to evaluate the performance of POD controllers.

The table of Figure 2 shows the data of the operating point. The power flow through line 2-3 is $P_{23} = 100$ MW. All data of the synthetic two-area test system can be found in [24].

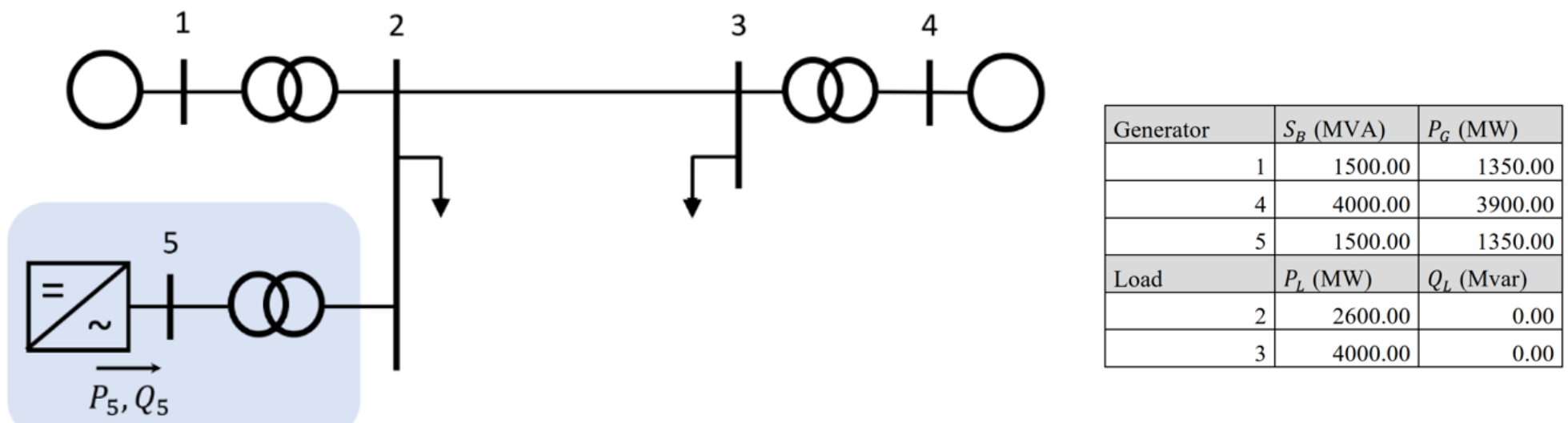

| Generator | $S_B$ (MVA) | $P_G$ (MW) |
|---|---|---|
| 1 | 1500.00 | 1350.00 |
| 4 | 4000.00 | 3900.00 |
| 5 | 1500.00 | 1350.00 |
| Load | $P_L$ (MW) | $Q_L$ (Mvar) |
| 2 | 2600.00 | 0.00 |
| 3 | 4000.00 | 0.00 |

*Figure 2: Synthetic two-area system (NTS).*

Relevant cases for the evaluation of the performance of POD controllers are the following:

- B0: Base case, only synchronous machines connected and without GFL device (in this case, $P_{L2} = 1250$ MW to achieve $P_{23} = 100$ MW).
- B1: GFL device connected, without POD controllers.
- B2: GFL device connected, with POD-P controller.
- B3: GFL device connected, with POD-Q controller.
- B4: GFL device connected, with POD-P and POD-Q controllers simultaneously (POD-PQ).

The performance of POD controllers includes the robustness and the effectiveness. The robustness of a POD controller means that it should have a reasonable behaviour to damp electromechanical oscillations within a certain frequency range (e.g., 0.1-2.5 Hz). The effectiveness of a POD controller means that it should damp significantly a target electromechanical oscillation (an electromechanical oscillation in a narrower frequency range, for example 0.1-0.3 Hz).

The guide for the implementation of power oscillation damping controllers [18] proposed the following acceptance criteria for POD controllers, using the synthetic test system of Figure 2 and the methodology of Spanish NTS standard [24]:

- **Acceptance criterion 1:** Acceptance criterion for the **robustness of the POD controller**. It evaluates the behaviour of the POD controllers in a range of frequencies of the electromechanical mode, following the methodology of NTS standard.

  Acceptance criterion: When POD-P, POD-Q and POD-PQ controllers are activated, the damping ratio of the electromechanical modes ($\varsigma_i$) shall be $\varsigma_i \geq 5$ %, for values of the line reactance within the range $X_L = 0.01 - 0.6$ pu.

- **<u>Acceptance criterion 2:</u>** Acceptance criterion for the **effectiveness of the POD controller**. It evaluates the behaviour of the POD controller for target electromechanical oscillation of a specific frequency. In the case of the Iberian Peninsula and the Continental Europe power system, inter-area oscillations with frequency about 0.1-0.3 Hz are a major concern. Therefore, a line reactance of $X_L = 0.6$ pu is used in the synthetic system for this test.

  Acceptance criterion: When POD-P, POD-Q and POD-PQ controllers are activated, the increment of the damping ratio of the electromechanical mode, in comparison with the base case B1 ($\Delta\varsigma_i = \varsigma_i - \varsigma_{B1}$) shall be $\Delta\varsigma_i \geq 5$ %, for a line reactance of $X_L = 0.6$ pu.

The acceptance criteria for POD controllers could be evaluated by small-signal stability analysis or by non-linear time-domain simulation, using electromechanical-type models (also known as Root-Mean-Square (RMS) models), as described in [18], [24]. Notice that the increment of the damping ratio of the electromechanical mode in acceptance criterion for the effectiveness of POD controller is not excessively high, in order to achieve compliance criteria for POD controllers feasible from a practical point of view in contexts of massive integration of power converters with POD controllers.

It is important to highlight that the method to evaluate the performance of POD controllers and compliance criteria presented in this paper has not regulatory value, it is just part of the guidelines for good practices of the implementation of POD controllers proposed in [18].

It is also important to recall that this synthetic test system is used to evaluate the performance of POD controllers in GFL devices connected to the Iberian Peninsula power system. The synthetic test system could be calibrated, depending on the characteristics of the critical electromechanical oscillations present in the particular power system considered, or even different synthetic test systems could be used.

## 4 Results: Synthetic two-area system and compliance criteria

The synthetic two-area system of Figure 2 is considered now, to assess the performance of POD controllers, following the procedure described in Section 3. The power converter connected at bus 5 (VSC-5) has GFL control and it is controlled with constant active- and reactive-power injections. Power converter VSC-5 is rated to 1500 MVA for normalisation, as described in the methodology of [24]. Generic models of VSCs are used, which were developed in an R&D project [28]. VSC-5 is equipped with POD-P and POD-Q controllers, with the block diagrams showed in Figure 3. POD controllers are based on an extended implementation, which contains a low-pass filter, a wash-out filter, a lead/lag filter, a gain and a saturator. POD-P/POD-Q controllers use as input signal the frequency deviation at the connection point with respect to the nominal frequency. POD controllers were tuned using eigenvalue-sensitivity methods [6], [26], [9] and their parameters are shown in Table 1. Data of the synthetic two-area system can be found in [24]. Simulations were carried out using commercial tools for time-domain simulation and small-signal stability analysis of power systems using RMS models [29].

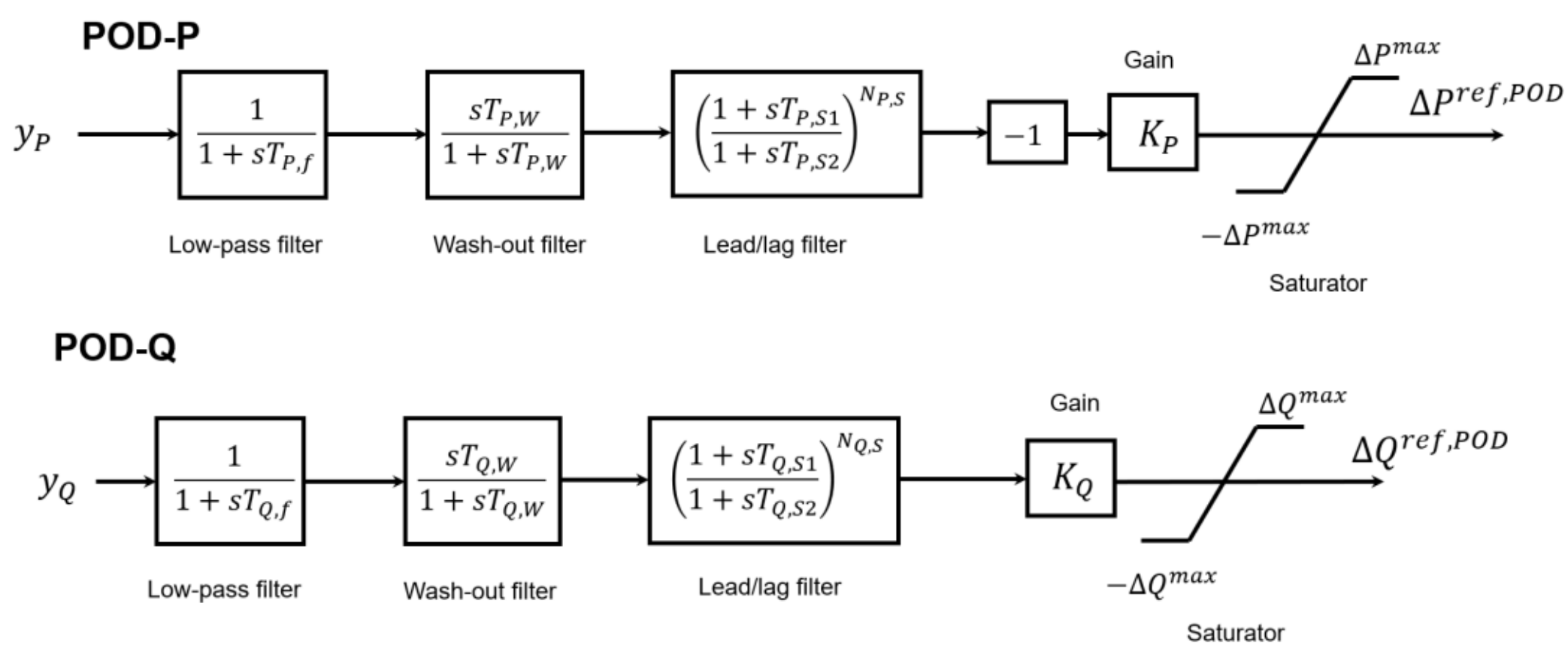

***Figure 3: POD controllers of VSC-5.***

***Table 1: POD controllers. Parameters.***

| POD-P | | POD-Q | |
|---|---|---|---|
| **Input: $y_P$** | Frequency deviation: $\Delta\omega = \omega - \omega_0$ (pu) | **Input $y_Q$** | Frequency deviation: $\Delta\omega = \omega - \omega_0$ (pu) |
| $K_P$ | 200 pu | $K_Q$ | 200 pu |
| $T_{P,S1}$ | 0.94 s | $T_{Q,S1}$ | 0.49 s |
| $T_{P,S2}$ | 1.07 s | $T_{Q,S2}$ | 1.02 s |
| $N_{P,S}$ | 2 | $N_{Q,S}$ | 2 |
| $T_{P,f}$ | 0.1 s | $T_{Q,f}$ | 0.1 s |
| $T_{P,W}$ | 5 s | $T_{Q,W}$ | 5 s |
| $\pm\Delta P^{max}$ | $\pm$0.1 pu | $\pm\Delta Q^{max}$ | $\pm$0.1 pu |

## 4.1 Scenario 1: Synthetic test system with generator G1 – PSS ON

In Scenario 1, generator G1 of the synthetic test system of Figure 2 has its PSS activated. Scenario 1 is the one used to evaluate the performance of POD controllers (compliance criteria) in [18], as described in Section 3.

### Small-signal stability analysis

Figure 4 shows the evolution of the system modes as the reactance of line 2–3 changes: $X_L = X_{23}$ from Figure 2. In case B1, VSC-5 is controlled with a constant reactive power setpoint, and its POD controllers are deactivated. POD-P/POD-Q controllers (cases B2, B3 and B4) prove to be effective, and the damping ratio of the electromechanical mode is improved in comparison to base case B1: the electromechanical mode shifts to the left in the complex plane (Figure 4).

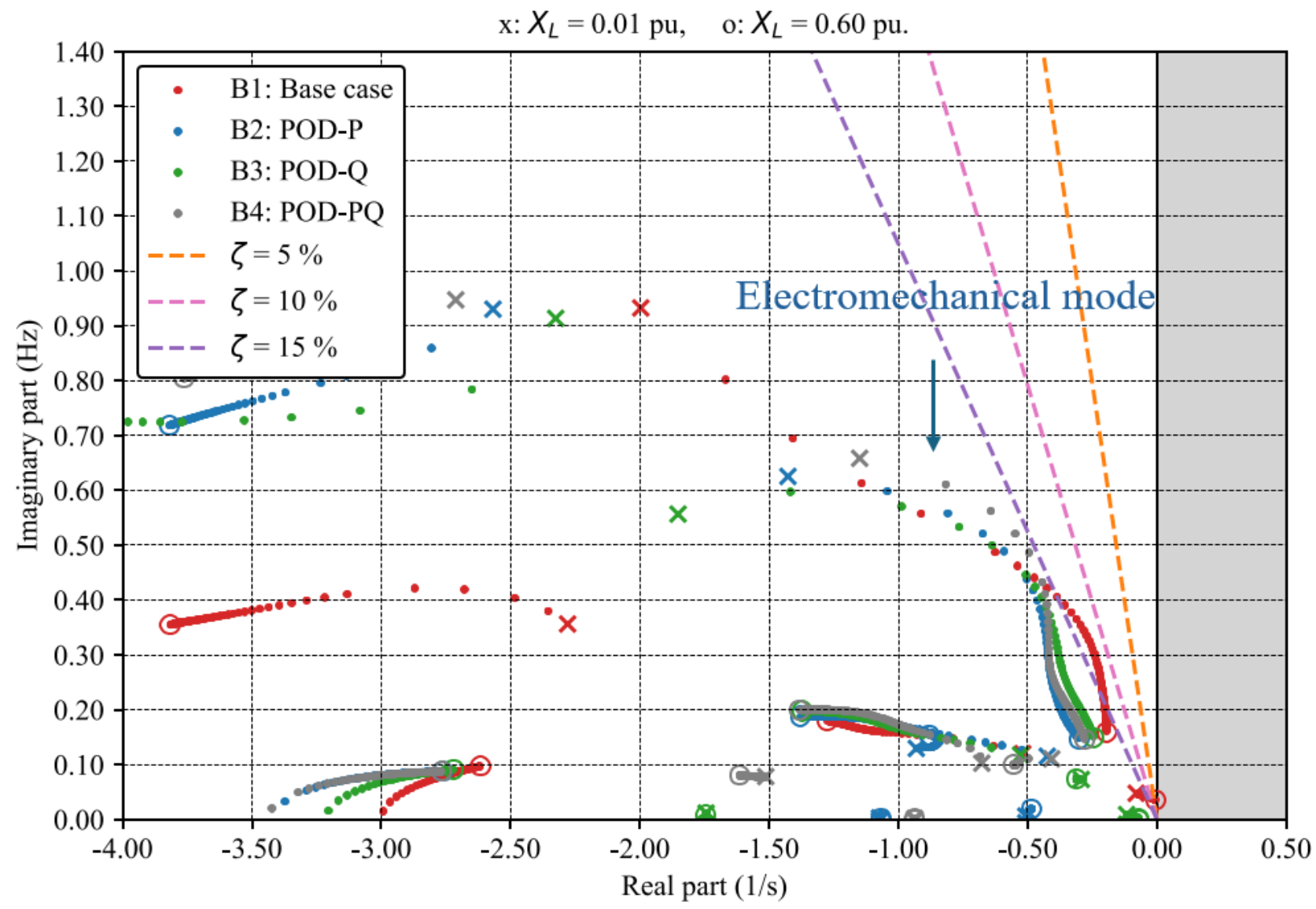

***Figure 4:Synthetic two-area test system - Scenario 1. Evolution of the eigenvalues of the system.***

Table 2 shows the damping ratio and oscillation frequency of the electromechanical mode under different operating modes for the four cases analysed. Previous conclusions are confirmed, the POD-P and POD-Q controllers of VSC-5 contribute to damp the electromechanical oscillation. POD-P/POD-Q controllers comply with the acceptance criterion for the robustness of the POD controllers, because the damping ratio of the electromechanical mode is higher than 5 % for all the values of the line reactance (electromechanical oscillation within the frequency range 0.1-1 Hz).

Table 3 quantifies the increment of the damping ratio of the electromechanical mode (for $X_L =$ 0.6 pu) obtained in each case. POD-P/POD-Q controllers comply with the acceptance criterion for the effectiveness of the POD controllers, because the increment of the damping ratio of the electromechanical mode is higher than 5 %.

***Table 2: Synthetic two-area test system - Scenario 1. Electromechanical modes.***

| | **B1: Base case** | | **B2: POD-P** | | **B3: POD-Q** | | **B4: POD-PQ** | |
|---|---|---|---|---|---|---|---|---|
| **$X_L$ (pu)** | ς (%) | $f_{mode}$ (Hz) | ς (%) | $f_{mode}$ (Hz) | ς (%) | $f_{mode}$ (Hz) | ς (%) | $f_{mode}$ (Hz) |
| **0.01** | 32.26 | 0.93 | 40.23 | 0.93 | 37.56 | 0.91 | 41.47 | 0.95 |
| **0.10** | 15.83 | 0.42 | 18.43 | 0.38 | 17.36 | 0.39 | 17.62 | 0.38 |
| **0.40** | 13.55 | 0.25 | 26.64 | 0.21 | 22.43 | 0.22 | 25.53 | 0.21 |
| **0.60** | 19.17 | 0.16 | 31.19 | 0.15 | 25.22 | 0.15 | 29.18 | 0.15 |

***Table 3: Synthetic two-area test system - Scenario 1. Damping ratio of the electromechanical mode for $X_L = 0.6$ pu.***

| **Case** | **$ς_i$ (%)** | **$ς_{B1}$ (%)** | **$\Delta ς_i$ (%)** |
|---|---|---|---|
| **B2: POD-P** | 31.19 | 19.17 | 12.02 |
| **B3: POD-Q** | 25.22 | 19.17 | 6.05 |
| **B4: POD-PQ** | 29.18 | 19.17 | 10.01 |

### Time-domain simulation

Results are checked now by non-linear time-domain simulation when the line reactance is $X_L = 0.6$ pu, which corresponds to an inter-area mode of 0.15 Hz approximately. A small disturbance is simulated to excite the electromechanical oscillation mode: a three-phase fault with high impedance at bus 2 (Figure 2), cleared after 50 ms. Figure 5-(a) shows the difference between the speeds of synchronous generators G1 and G4 of the synthetic two-area test system (Figure 2). This variable is useful to observe inter-area oscillations. Figure 5-(b) shows the active- and reactive-power injections of VSC-5.
Results confirm that the damping ratio of the electromechanical mode in the case with POD-P/POD-Q controllers (cases B2, B3 and B4) is greater than the one obtained in the base case B1. The damping ratio of the electromechanical mode is increased by means of active- and reactive-power modulation of POD-P/POD-Q controllers (see Figure 5-(b)).
In the base case B1 (without POD controller), the damping ratio of the electromechanical mode is low. In contrast, in case B2, once the POD controller is activated, the damping ratio of the electro-mechanical mode increases, confirming that it is an effective control strategy.

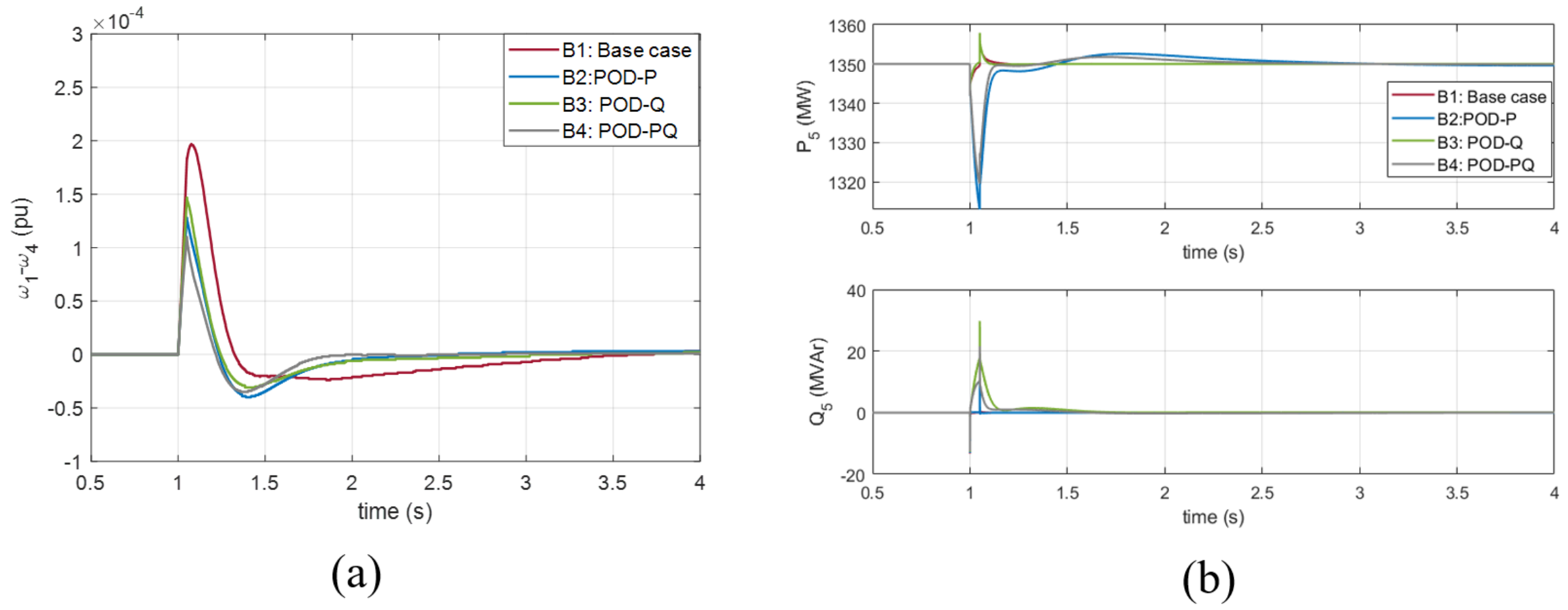

*Figure 5: Synthetic two-area test system - Scenario 1. (a) Difference between the speeds of synchronous generators G1 and G4; (b) Active- and reactive-power injections of VSC-5.*

## 4.2 Scenario 2: Generator G2 – PSS OFF

In Scenario 2, generator G1 of the synthetic test system of Figure 2 has its PSS deactivated. As explained in previous subsection, Scenario 1 is the one used to quantify compliance criteria of POD controllers in the guide [18]. Scenario 2 is just used to analyse the performance of POD controllers in scenarios in which the electromechanical oscillations have a low damping ratio, or even negative. This could be useful to analyse with further details the performance of POD controllers.

### Small-signal stability analysis

Figure 6 shows the evolution of the system modes as the reactance of line 2–3 changes: $X_L = X_{23}$ from Figure 2. In case B1 (POD controllers OFF), the system is unstable for some values of the line reactance, because the electromechanical mode has positive real part those values of the line reactance, which correspond to low-frequency inter-area oscillations. POD-P/POD-Q controllers (cases B2, B3 and B4) prove to be effective, and the damping ratio of the electromechanical mode is significantly improved in comparison to base case B1: the electromechanical mode shifts to the left in the complex plane (Figure 6). With POD controllers, the electromechanical oscillation has positive damping ratio for all values of the line reactance. Notice that the increments of the damping ratio of the electromechanical mode in Scenario 2 are higher than the ones obtained in Scenario 1 (see Table 4 for $X_L = 0.6$ pu). This is because, typically, the improvements of POD controllers are higher when the base case (B1) is more critical.

***Table 4: Synthetic two-area test system - Scenario 2. Damping ratio of the electromechanical mode for $X_L = 0.6\ pu$.***

| Case | $\varsigma_i$ (%) | $\varsigma_{B1}$ (%) | $\Delta\varsigma_i$ (%) |
|---|---|---|---|
| **B2: POD-P** | 32.61 | -30.75 | 63.36 |
| **B3: POD-Q** | 26.02 | -30.75 | 56.77 |
| **B4: POD-PQ** | 30.86 | -30.75 | 61.61 |

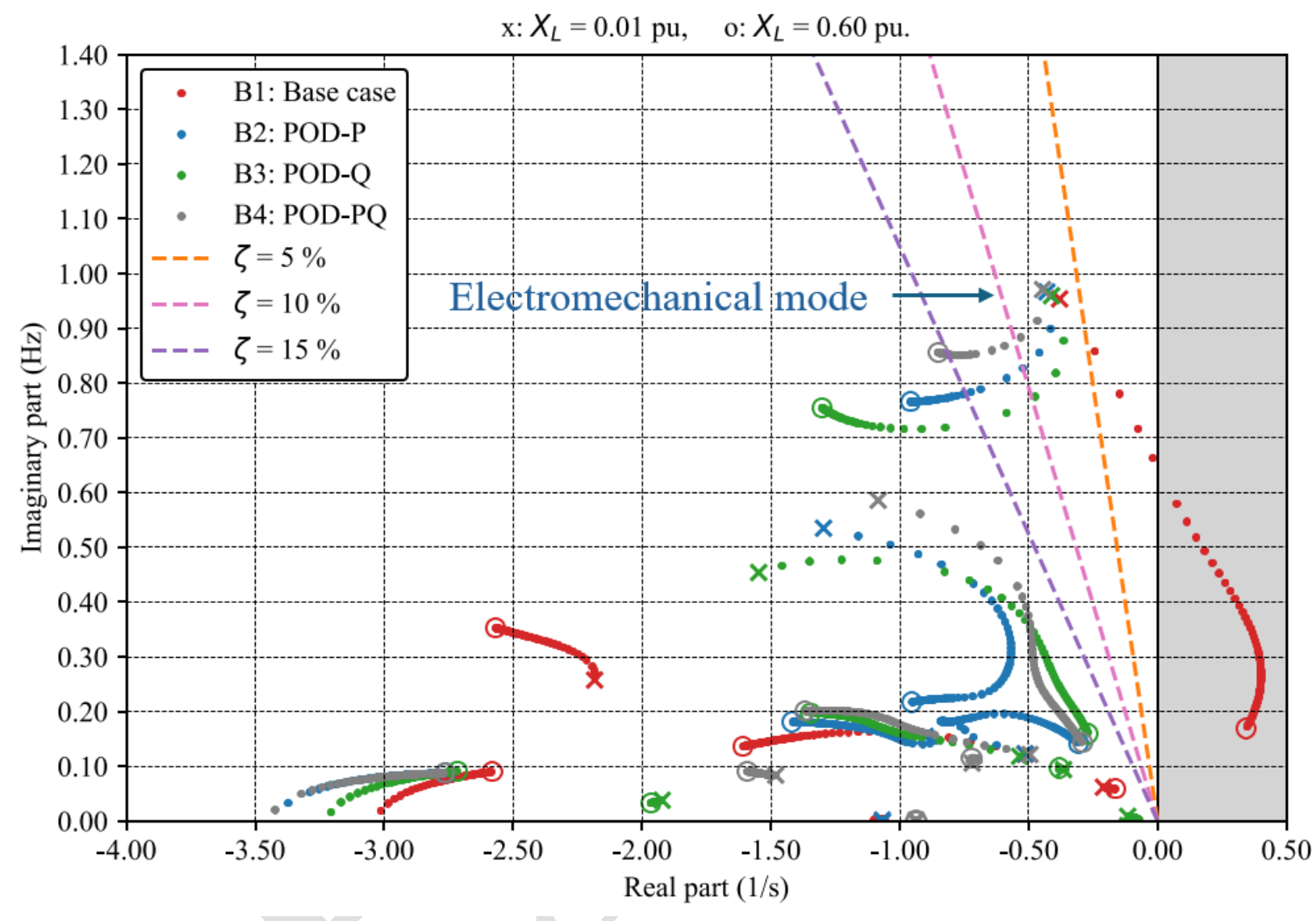

***Figure 6: Synthetic two-area test system - Scenario 2. Evolution of the eigenvalues of the system.***

## Time-domain simulation

Results are checked now by non-linear time-domain simulation when the line reactance is $X_L = 0.6$ pu, which corresponds to an inter-area mode of 0.15 Hz approximately. Again, a small disturbance was simulated: a three-phase fault with high impedance at bus 2 (Figure 2), cleared after 50 ms. Figure 7-(a) shows the difference between the speeds of synchronous generators G1 and G4 of the synthetic two-area test system (Figure 2).
Results confirm that base case B1 (with POD controllers deactivated) is unstable, while the damping ratio of the electromechanical mode in the case with POD-P/POD-Q controllers (cases B2, B3 and B4) is positive and greater than the one obtained in the base case B1. The electromechanical oscillation was damped by means of active- and reactive-power modulation of POD-P/POD-Q controllers (Figure 7-(b)).

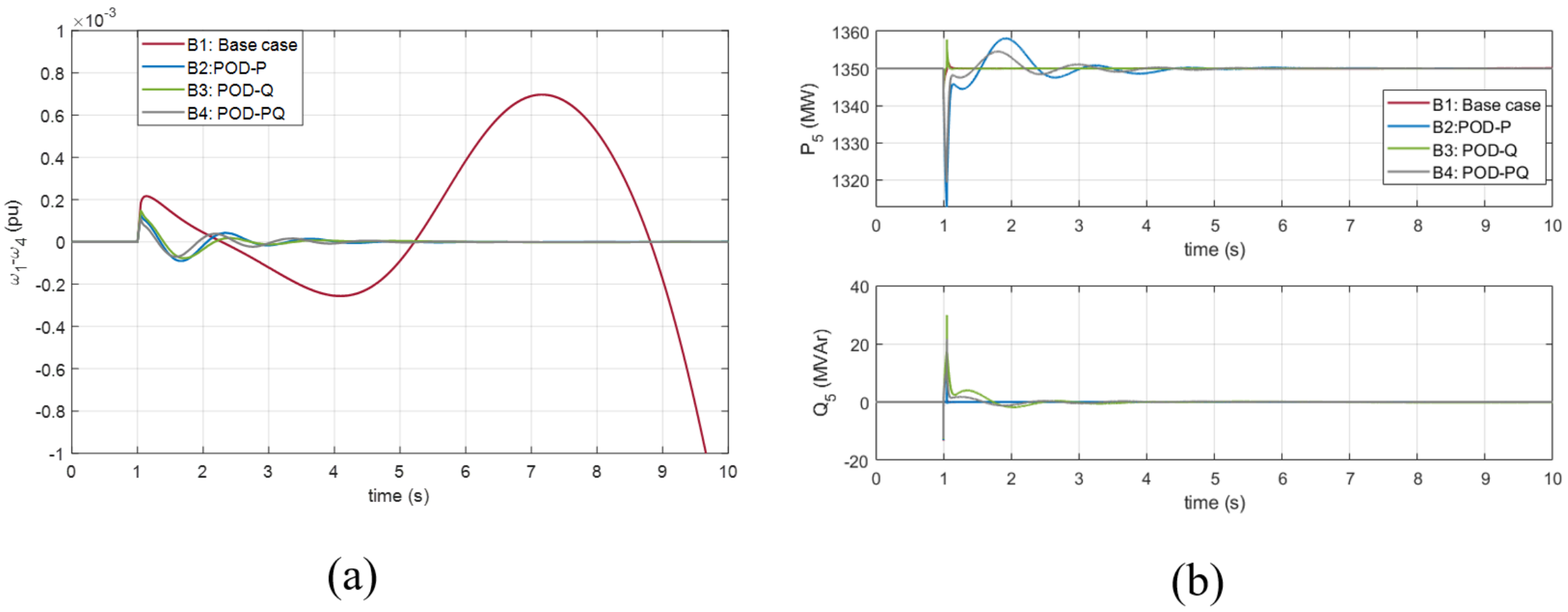

***Figure 7: Synthetic two-area test system - Scenario 2. (a) Difference between the speeds of synchronous generators G1 and G4; (b) Active- and reactive-power injections of VSC-5.***

## 5 Case Study: Large-scale power system

A large-scale power system is considered now, in order to validate the effectiveness of POD controllers tuned in the synthetic system of Figure 1 and using the compliance criteria described in Section 3. The objective is to check if the POD controller also produces satisfactory results when the power converter is connected to a large-scale power system.

A detailed model of the Iberian Peninsula power system is considered, connected to a detailed model of the rest of Continental Europe power system [30], [31]. The scenario considered does not represent any realistic scenario: the model was calibrated for this specific study, in order to obtain an East-Centre-West inter-area oscillation with low damping ratio. Scripts and models developed in an R&D project were used during the process of calibrating this scenario [32].

Two VSC converters with GFL control located in arbitrary buses in the Iberian Peninsula power system are considered:

- VSC-1 (200 MVA): Located in the southern area of the Iberian Peninsula power system (Carmona 400 kV).
- VSC-2 (200 MVA): Located in the central area of the Iberian Peninsula power system (La Cereal 400 kV).

These two locations were selected just for illustration purposes in this study and just because they have Phasor Measurement Units (PMU) and their frequencies are often monitored. The rating of the VSCs has also been selected arbitrarily (200 MVA). The same generic models of the power converters than in Section 7 were used and with the same parameters of POD-P/POD-Q controllers (Table 1). Only the rating of the power converters was changed. The GFL VSCs are controlled with constant active- and reactive-power injections, with zero P/Q power injections at the initial operating point (P=Q=0). Small-signal stability analysis was carried out using a commercial tool with RMS models [29].

Table 5 shows the damping ratio and the frequency of the East-Centre-West inter-area oscillation. Results show that POD-P/POD-Q controllers of VSC-1 and VSC-2 contribute to damp the inter-area oscillation. These POD controllers were tuned used the synthetic system of Figure 1 and using the compliance criteria described in Section 3.

***Table 5: Large-scale power system. Damping ratio of the critical inter-area mode. POD-PQ: both POD-P and POD-Q activated.***

| Case | $\varsigma$ (%) | $f_{mode}$ (Hz) |
|---|---|---|
| **Base case: PODs OFF in VSC-1 and VSC-2** | 1.16 | 0.20 |
| **VSC-1: POD-P** | 2.56 | 0.24 |
| **VSC-1: POD-Q** | 1.36 | 0.20 |
| **VSC-1: POD-PQ** | 2.64 | 0.24 |
| **VSC-2: POD-P** | 2.18 | 0.24 |
| **VSC-2: POD-Q** | 1.39 | 0.20 |
| **VSC-2: POD-PQ** | 2.26 | 0.24 |
| **VSC-1 & VSC-2: POD-PQ** | 3.17 | 0.24 |

Finally, results are checked by non-linear time-domain simulation. Only two cases are compared: the base case with POD-PQ OFF versus the case with both POD-PQ ON in VSC-1 and VSC-2. A load of 594 MW / 229 MVAr in the CE power system is disconnected at $t = 1$ s. Figure 8-(a) shows the difference between the speeds of a synchronous generator in Spain (G-A) and a synchronous generator in Germany (G-B) and the frequencies at the AC connection point of the VSCs. Results show that POD-P/POD-Q controllers of VSC-1 and VSC-2 contribute to damp the inter-area oscillation, due to P/Q modulation of the VSCs (Figure 8-(b)).

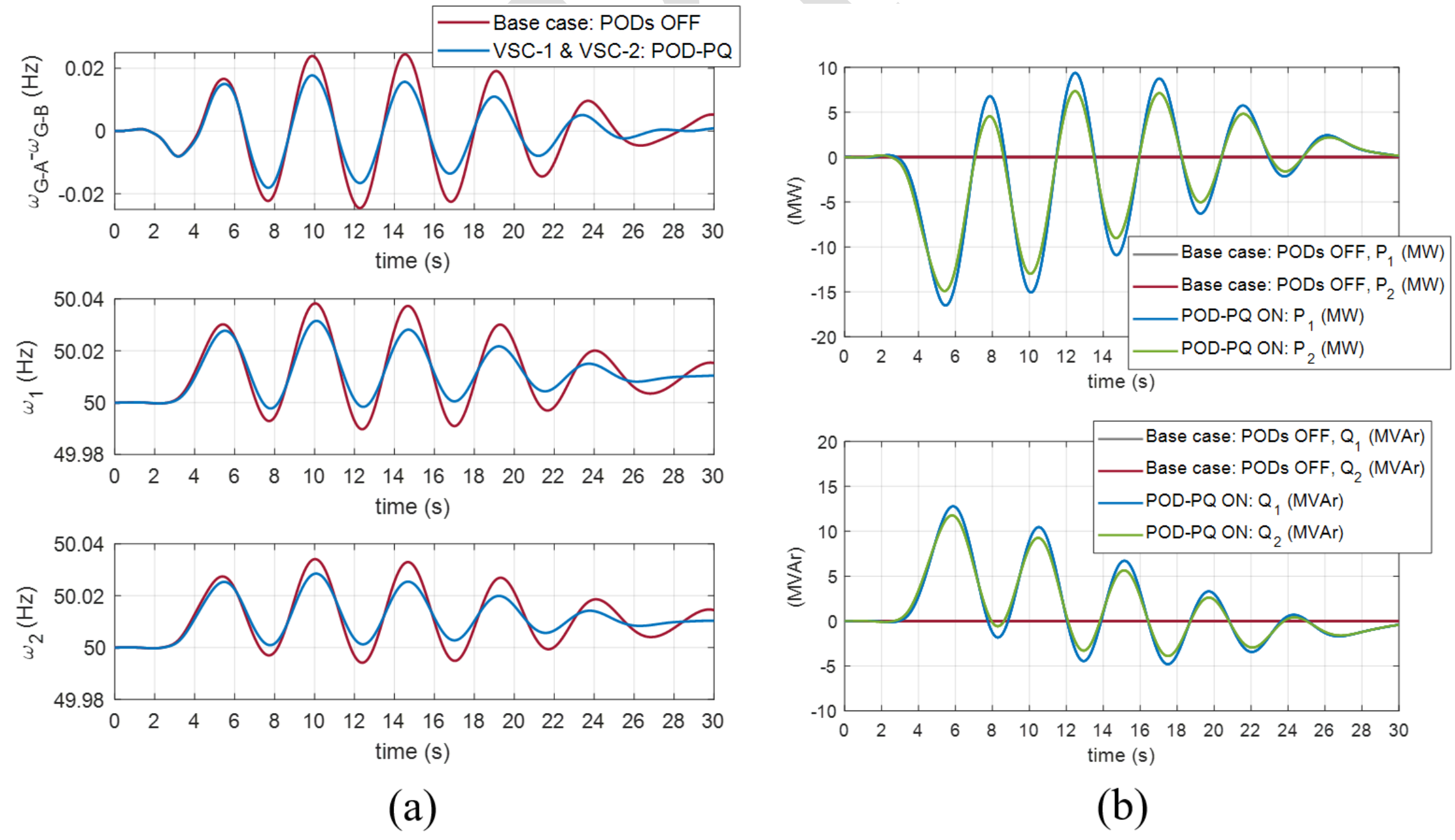

***Figure 8: Large-scale power system 1. (a) (top) Difference between the speeds of synchronous generators G-A (Spain) and G-B (Germany), (middle) frequency at the AC bus of VSC-1 and (bottom) frequency at the AC bus of VSC-2. (b) (top) Active-power injections and (bottom) reactive-power injections of the VSCs.***

Hence, results prove that by using appropriate synthetic systems, methodologies and compliance criteria, POD controllers could be useful to damp electromechanical oscillation in large-scale power systems.

## 6 Conclusions

This paper provided guidelines for the implementation of POD controllers in power converters. The paper proposed compliance criteria for POD controllers using a synthetic test system and a systematic methodology used in Spanish technical standard for monitoring compliance (NTS), considering practical aspects. The paper also included numerical examples to illustrate compliance criteria for POD controllers in a synthetic test system. A generic power converter with GFL is used for the analysis by simulation and POD controllers using modulation of active-power injection (POD-P), reactive-power injection (POD-Q) or both simultaneously (POD-PQ) were analysed. Results were validated in a large-scale power system.

The paper concludes that by using appropriate synthetic systems, methodologies and compliance criteria, POD controllers in power converters could be effective to damp electromechanical oscillation in large-scale power systems. This result has a great interest from a practical point of view, especially for tuning POD controllers in power converters when limited information of the power system is available.